\documentclass{mn2e}
\usepackage{aas_macros,astrobib}
\usepackage[a4paper,centering, totalwidth=520pt, totalheight=700pt]{geometry}

\usepackage{graphicx}
\usepackage{amssymb}
\usepackage{amsmath}
\usepackage{enumerate}
\usepackage{microtype}

\usepackage{color}

\newcommand{\eqn}[1]{equation~\eqref{#1}}

\newcommand{\fig}[1]{Figure~\ref{#1}}

\newcommand{\be}{\begin{equation}}
\newcommand{\ee}{\end{equation}}
\newcommand{\bear}{\begin{eqnarray}}
\newcommand{\ear}{\end{eqnarray}}

\newcommand{\f}{\frac}

\newcommand{\der}{\ensuremath{{\rm d}}}

\newcommand{\subB}{\scriptscriptstyle \rm B}

\begin{document}

\title[Reionization and magnetic fields]
{Reionization constraints on primordial magnetic fields }
\author[Pandey, Choudhury, Sethi \& Ferrara]
{Kanhaiya L. Pandey$^1$\thanks{Email: kanhaiya@ncra.tifr.res.in}, T. Roy Choudhury$^1$, Shiv K. Sethi$^2$ and Andrea Ferrara$^3$
\\
$^1$National Centre for Radio Astrophysics, TIFR, Pune 411007, India\\
$^2$Raman Research Institute, Bangalore 560080, India\\
$^3$Scuola Normale Superiore, Piazza dei Cavalieri 7, I-56126 Pisa, Italy
} 

\maketitle

\date{\today}

\begin{abstract}
We study the impact of the extra density fluctuations induced by primordial magnetic fields on the reionization history in the redshift
range: $6 < z < 10$. We perform a comprehensive MCMC physical analysis allowing the variation of parameters related to primordial 
magnetic fields (strength, $B_0$, and power-spectrum index $n_{\subB}$), reionization, and $\Lambda$CDM cosmological model. We 
find that magnetic field strengths in the range: $B_0 \simeq 0.05\hbox{--}0.3$ nG (for nearly scale-free power spectra) can significantly 
alter the reionization history in the above redshift range and can relieve the tension between the WMAP and quasar absorption 
spectra data. Our analysis puts upper-limits on the magnetic field strength $B_0 < 0.358, 0.120, 0.059$ nG (95 \% c.l.) for $n_{\subB} = 
-2.95, -2.9, -2.85$, respectively. These represent the strongest magnetic field constraints among those available from other cosmological 
observables.
\end{abstract}

\section{Introduction}

\label{outline}

Magnetic fields are ubiquitously present in the universe and play 
an important role in various astrophysical processes:  
star formation, accretion disks, proto-planetary disks, formation 
and stability of jets, dynamics of inter stellar medium (ISM), etc. (e.g. 
\citeNP{1979cmft.book.....P}, \citeNP{1983flma....3.....Z})
However their role is still not well understood on  larger
scales and in the cosmological context, e.g.  in the process 
of formation of the early galaxies and structure formation (e.g.
\citeNP{2002RvMP...74..775W}, \citeNP{2012SSRv..166....1R}). 
At present,  the possible impact of  cosmic magnetic fields is being 
investigated for a host of observables in  astrophysics and cosmology. 
The probe of such fields is also one of the principle aims of 
many of the ongoing and 
upcoming large radio interferometers  such as LOFAR\footnote{\tt http://www.lofar.org/} and  SKA\footnote{\tt https://www.skatelescope.org/}.

Magnetic fields in the universe are known to  be coherent over the scales
of galaxies and galaxy clusters, $\simeq 10\hbox{--}50$~kpc (e.g. 
\citeNP{2012SSRv..166..215B}, \citeNP{2002RvMP...74..775W}). There is also 
evidence of  coherent magnetic fields  over super-cluster scales 
(\citeNP{1989Natur.341..720K}).  These 
 fields could have arisen from the dynamo amplification of 
very small seed fields ($\simeq 10^{-20}\, \rm G$) generated in the early universe (e.g. \citeNP{2012SSRv..166..215B}).
Alternatively, the observed fields might owe their origin 
to far stronger large scale primordial fields   ($\sim 10^{-9} G$) 
that  could have been generated during inflation or 
other an early phase transition in the universe (\citeNP{1988PhRvD..37.2743T}, 
\citeNP{1992ApJ...391L...1R}, \citeNP{2002RvMP...74..775W}, \citeNP{2005PhRvD..71j3509A}). 
Here we study some implications of the 
latter hypothesis. 

Primordial magnetic fields leave their signature on a  range of cosmological
processes and observables (e.g. \citeNP{2012PhR...517..141Y}, \citeNP{2014PhRvD..89h3528Y}). These fields
can  generate density perturbations in addition to the 
$\Lambda$CDM model in  the 
post-recombination epoch  (\citeNP{1978ApJ...224..337W}, 
\citeNP{1996ApJ...468...28K}, \citeNP{2003JApA...24...51G}). The matter power
 spectrum of  these density perturbations  
dominates the standard $\Lambda$CDM matter power spectrum at small scales 
($\simeq 1 \hbox{--} 10 \, \rm h^{-1} \,  Mpc$). Such effects can be directly probed by cosmological gravitational lensing and Lyman-$\alpha$ observations
(\citeNP{2014MNRAS.437.3639C}, \citeNP{2013ApJ...762...15P}, \citeNP{2012ApJ...748...27P}, \citeNP{2012PhRvD..86d3510S}). 

This additional power can cause early formation of structures (e.g., galaxies) which consequently cause early reionization of the IGM. As of today, the main observational constraints on reionization come from the CMB polarization data (e.g., those from the WMAP and Planck experiments, \citeNP{2013arXiv1303.5062P}) and the quasar absorption spectra at $z \gtrsim 5.5$ (\citeNP{2006AJ....132..117F}). Detailed models which are able to match these and a variety of other observations predict the reionization to be process extended over $6 \lesssim z \lesssim 15$. It has been shown that the presence of magnetic fields can affect the reionization history and alter the HI signal from epoch of reionization (\citeNP{2012PhR...517..141Y}, \citeNP{2009JCAP...11..021S}, \citeNP{2010PhRvD..82h3005K}, \citeNP{2005MNRAS.356..778S}). In this paper, we extend available detailed models of reionization (\citeNP{2005MNRAS.361..577C}, \citeNP{2006MNRAS.371L..55C}) by including magnetic fields and carry out a detailed mutli-parameter MCMC analysis to compare with available data sets. The main goal is to check the level of constraints one can put on the primordial magnetic field.

There is another aspect related to the magnetic fields, early formation of galaxies and reionization which is worth considering. Models which are consistent with all available data sets usually require some efficient sources of ionizing photons at high redshifts, e.g., metal-free PopIII stars \cite{2005MNRAS.361..577C,2011MNRAS.413.1569M} in addition to the usual PopII stars. These PopIII sources, however, are unlikely to contribute significantly to the photon budget at lower redshifts $z \lesssim 9$ because of feedback effects, which is crucial in matching the low photoionization rate inferred from quasar absorption spectra $z \sim 6$ \cite{2007MNRAS.382..325B}. One can thus conclude that in order to match the data sets, one requires a high ionizing emissivity at $z \sim 10$ which should preferably decrease by $z \sim 6$. In this regard, the presence of magnetic fields may assist in producing large number of dark matter haloes at high redshifts, and hence would help in reconciling with data sets without invoking any other sources like PopIII stars.
This paper considers the 
effect of primordial magnetic field on the structure formation as a viable 
source which can affect the reionization process appreciably.

The plan of the paper is as follows: we discuss the model of reionization and 
the impact of the inclusion of primordial  magnetic field in the next section. The main results are presented in Section~3. We summarize our results and discuss their implications in section~4. 

\section{Method}
We first briefly summarize the main features of the reionization model used in this paper; the details can be found in \citeN{2005MNRAS.361..577C}, \citeN{2006MNRAS.371L..55C} and  \citeN{2009CSci...97..841C}. The model follows the ionization and thermal histories of
neutral, HII and HeIII regions simultaneously and self-consistently accounting
for the IGM inhomogeneities based on a method outlined in
\citeN{2000ApJ...530....1M}. The density distribution of the IGM is
assumed to be lognormal. The rate of number of ionizing photons
per unit volume (i.e., ionizing emissivity)
in the IGM from galaxies is assumed to be given by
\be
\dot{n}_{\rm ph}(z) = n_b N_{\rm ion} \f{\der f_{\rm coll}}{\der t},
\label{eq:ndot}
\ee
where $n_b$ is the number density of baryons and $f_{\rm coll}$ is the
dark matter collapsed fraction. 
The proportionality constant
$N_{\rm ion}$ physically represents the number of ionizing photons
in the IGM per baryon in collapsed objects. It can be written as
\be
N_{\rm ion} = \epsilon_* f_{\rm esc} m_p \int_{\nu_{\rm HI}}^{\infty} \der \nu \left[\f{\der N_{\nu}}{\der M_*}\right],
\ee
where $\epsilon_*$ is the fraction of baryons within collapsed haloes
going into stars, $f_{\rm esc}$ is the escape fraction of ionizing
photons and $[\der N_{\nu}/\der M_*]$ gives the number of ionizing photons
(i.e., those with frequencies higher than the ionization
threshold $\nu_{\rm HI}$)
per frequency interval per unit stellar mass and is determined by the
stellar IMF and the corresponding stellar spectra, $m_p$ is proton-mass. We assume the 
stars to be Population II with subsolar metallicity $Z = 0.001 = 0.05 Z_{\odot}$ with a
Salpeter IMF in the mass range $1 - 100 M_{\odot}$. With this
assumption, the parameter $N_{\rm ion}$ will be given by
\be
N_{\rm ion} \approx 3200 \epsilon_{*, {\rm II}} f_{\rm esc, II} 
= 3200 \epsilon_{\rm II},
\ee
where we have defined $\epsilon_{\rm II} \equiv \epsilon_{*, {\rm II}} f_{\rm esc, II}$.
We assume $\epsilon_{\rm II}$ to be independent of redshift and consider it as a free parameter in our model. In this work, we do not invoke any other sources of reionization which are often done in other studies, e.g., metal-free PopIII stars. Our main aim would be to verify if the presence of magnetic fields is able to eliminate the requirement of PopIII stars and still give a good match to the data.

The collapse fraction, $f_{\rm coll}$, depends on the minimum mass of star-forming haloes. In neutral
regions, we assume it to be determined by atomic cooling (i.e. we neglect
cooling through molecular hydrogen). However, the minimum mass
will be larger in ionized regions because of radiative feedback. Our
model can compute radiative feedback (suppressing star formation in
low-mass haloes using a Jeans mass prescription) self-consistently
from the evolution of the thermal properties of the IGM. 
The corresponding filtering scale, which depends on the 
temperature evolution of the IGM, is found to be typically around 
$\sim 30$ km s$^{-1}$.

The HI photoionization rate is given by
\be
\Gamma_{\rm HI}(z) =4 \pi (1+z)^3 \int_{\nu_{\rm HI}} \der \nu
~\lambda_{\rm mfp}(z,\nu)~
\dot{n}_{z,\nu}~\sigma_{\rm HI}(\nu),
\ee
where $\sigma_{\rm HI}$ is the photoionization cross section and
$\lambda_{\rm mfp}$ is the mean free path of ionizing photons. The mean
free path is modelled as \cite{2000ApJ...530....1M}
\be
\lambda_{\rm mfp}(z) = \f{\lambda_0}{[1 - F_V(z)]^{2/3}},
\ee
where $F_V$ is the volume fraction of ionized regions and 
$\lambda_0$ is a normalization parameter. This parameter can be constrained
using the redshift distribution of Lyman-limit systems
\be
\f{\der N_{\rm LL}}{\der z} = \f{c}{\sqrt{\pi} \lambda_{\rm mfp}(z) H(z) (1+z)}.
\ee
Given a reionization history, we compute the angular power spectra $C_l$
of CMB temperature and ($E$-mode) polarization anisotropies. We combine
our calculations with the 
publicly available code CAMB\footnote{\tt http://camb.info/} \cite{2000ApJ...538..473L} in order to do so. The crucial
parameter which determines the $C_l$ is the electron scattering
optical depth
\be
\tau_{\rm el}(z) = \sigma_T c \int_0^{t(z)} \der t~n_e(z)~(1+z)^3,
\ee
where $n_e(z)$ is the comoving number density of free electrons and
$\sigma_T$ is the Thomson cross section and $c$ is the speed of light.

\subsection{Inclusion of magnetic fields}

The effect of the magnetic field is included by adding the additional
matter  power induced by these fields 
to the usual dark matter power spectrum $P_{\rm DM}(k)$. 
As we show below, the collapsed fraction  $f_{\rm coll}$, which determines the
ionizing emissivity in \eqn{eq:ndot}, is a 
sensitive function of $P_{\rm DM}(k)$ and hence the inclusion of magnetic fields
can significantly alter  the reionization history.

We assume the  primordial magnetic field to  be 
a  stochastic Gaussian field (see \citeNP{2002PhRvD..65l3004M} and references therein). 
For a  non-helical magnetic field, the two-point correlation 
function  of the tangled 
field can be written as: 
\begin{equation}
 \langle B^\star_i({\mathbf k})B_j({\mathbf k'})\rangle
=(2\pi)^3 \delta^{(3)} ({\mathbf k}-{\mathbf k'})
P_{ij}({\mathbf{\hat k}}) P_{\subB}(k). \label{spectrum}
\end{equation}
Here $i$ and $j$ are spatial indices, $i,j \in (1,2,3)$,
$\hat{k}_i=k_i/k$  a unit wave vector, $P_{ij}({\mathbf{\hat
k}})=\delta_{ij}-\hat{k}_i\hat{k}_j$
 the transverse plane projector,
 $\delta^{(3)}({\mathbf k}-{\mathbf k'})$  the Dirac delta
 function, and $P_{\subB}(k) = Ak^{n_{\subB}}$ is the power spectrum of the magnetic
field; here $A$ normalizes the power in magnetic fields and $n_{\subB}$ is the  
spectral index. 
The parameter $A$ is computed by defining the RMS of magnetic field
at cut off scale $k_c$; the RMS for  $k_c = 1 \, \rm Mpc^{-1}$, 
is referred to as the magnetic field strength, $B_0$  (\citeNP{2002PhRvD..65l3004M}). The relation between $A$ and $B_0$ is given by
\begin{equation}
A = \frac{(2\pi)^{n_{\subB}+5} B^2_0}{2\Gamma(n_{\subB}/2+3/2)}.
\label{energy-spectrum-H}
\end{equation}

The magnetic field power spectrum drops at small scales 
owing to dissipation in the pre-recombination era. 
The cut-off scale $k_{\rm max}$ is determined by the Alfv\'en wave
damping scale $k_{\rm max}^{-1} \sim v_A L_S$ where $v_A$ is the Alfv\'en
velocity and $L_S$ the Silk damping scale (\citeNP{1998PhRvD..57.3264J}):
$k_{\rm max} \simeq 200 \, (1 \, {\rm nG}/B_0) \, \rm Mpc^{-1}$.

\begin{figure}
\begin{center}
\includegraphics[width=\columnwidth]{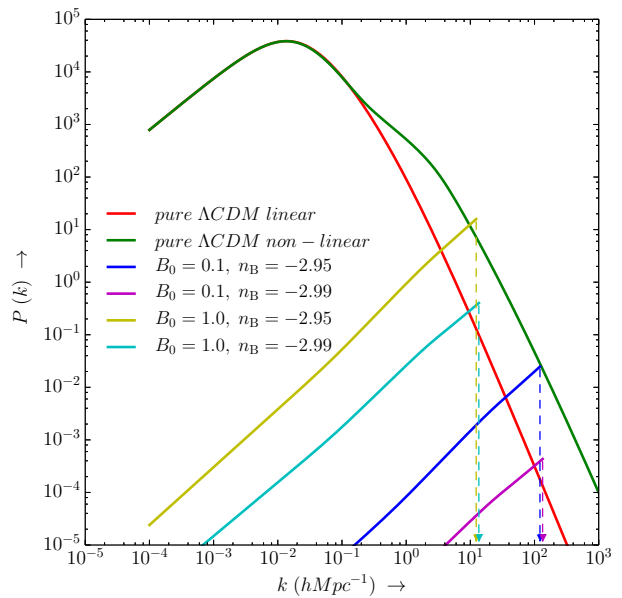}
\end{center}
\caption{Inflationary and primordial magnetic field induced 
matter power spectra. A sharp cut-off has been put 
at the magnetic Jeans scale for magnetic cases.} \label{fig:test}
\end{figure}

The primordial  magnetic field induces density perturbations in the
post-recombination era which grow by gravitational collapse
 (\citeNP{1978ApJ...224..337W}, \citeNP{1996ApJ...468...28K}, 
\citeNP{2003JApA...24...51G}).
The matter power spectrum induced by magnetic fields has the shape: 
 $P(k) \propto k^{2n_{\subB} +7}$ for $n_{\subB} \le -1.5$; this  matter power spectrum 
is  cut off at the magnetic field Jeans' scale: $k_J \simeq 15( 1\, {\rm nG}/B_{0}) \, \rm Mpc^{-1}$ (\citeNP{1996ApJ...468...28K}, \citeNP{2003JApA...24...51G})
\footnote{The quantity $k_{\rm max}$ is a cut-off scale in magnetic field power spectrum $P_{\subB}(k)$ in eq~\ref{spectrum}, 
whereas $k_J$ is a magnetic Jeans cut-off scale in magnetic field induced matter power spectrum due to 
magnetic pressure. The magnetic Jeans scale does depend on $n_{\subB}$, but the dependence is extremely weak for 
near scale-free models $n_{\subB} \simeq -3$.}. 
The matter power spectrum is shown in the \fig{fig:test} for different  values of $B_0$ and $n_{\subB}$ along with the power spectrum for the usual $\Lambda$CDM case. 
As shown in the Figure, we adopt a sharp cut-off of the matter 
power spectrum at the magnetic Jeans' scale  $k = k_J$ (\citeNP{1980lssu.book.....P,2005MNRAS.356..778S,2006MNRAS.368..965T,2006MNRAS.372.1060T,2009ApJ...703.1096S,2011MNRAS.411.1284T,2011MNRAS.418L.143S,2012MNRAS.424..927T,2012ApJ...748...27P}).  
We note that the mass
dispersion $\sigma^2 \propto k^3 P(k)$ and the number of haloes 
for  a given mass is very sensitive to $\sigma$.  Therefore, 
by  adopting a sharp cut-off, we obtain conservative  limits on 
the strength of  magnetic field as compared to the case in which 
the cut-off is  gradual \cite{2013ApJ...762...15P}.

It has been shown from other cosmological observables that the only class
of acceptable magnetic field models correspond to the near scale-free 
models, $n_{\subB} \simeq -3$ (e.g. \citeNP{2005MNRAS.356..778S}). 
Hence in this work we study only models with $n_{\subB}$ very close to $-3$.

In this paper  we assume the sources of inflationary density 
perturbations and magnetic field generation to be uncorrelated. This allows
us to add the two matter power spectra in quadrature for our computation. It
 should 
also be underlined that the presence of sub-nG fields does not substantially  change
the normalization of $\sigma_8$, as magnetic field induced matter power spectra
make negligible contribution to the scales of interest: $k \simeq 0.01\hbox{--}0.5 \, \rm Mpc^{-1}$ (\fig{fig:test}). 
As noted above, the main impact of the extra matter power induced by magnetic fields is
to increase the collapse fraction and alter its evolution. This induced 
power spectrum (\fig{fig:test}) causes collapse of mass haloes
close to the magnetic field Jeans' scale (e.g. \citeNP{2010PhRvD..82h3005K})
and therefore changes ionization history which might not be reproducible 
by altering parameters within the framework of $\Lambda$CDM model.

\begin{figure*}
\centering
\includegraphics[angle=0,width=0.99\textwidth]{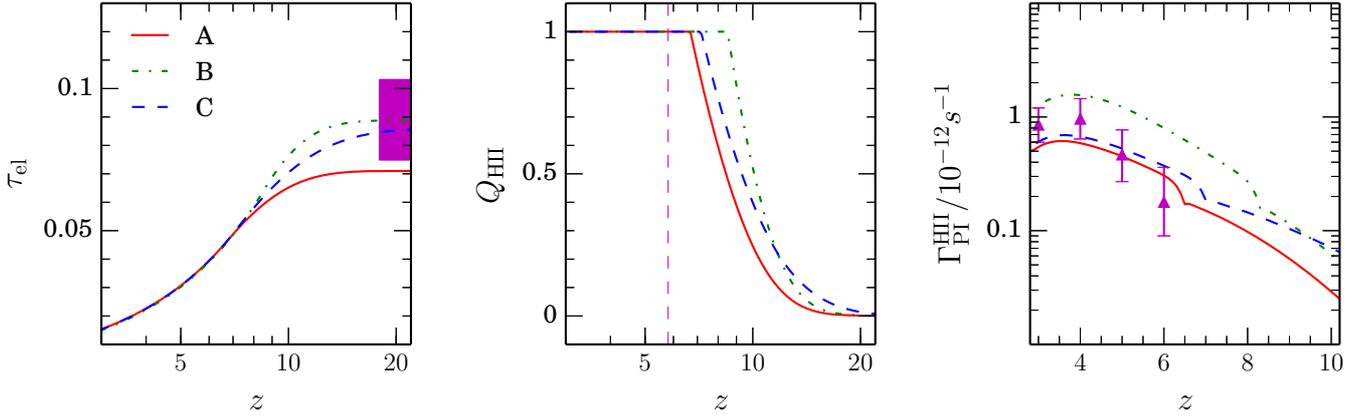}
\caption{Effect of non-zero $B_0$ on quantities related to reionization. The left hand panel shows the evolution of electron scattering optical depth $\tau_{\rm el}$, the middle panel shows the volume filling factor of ionized regions and the right hand panel shows the evolution of the photoionization rate $\Gamma_{\rm HI}$. We have also shown the observational data points, see text for details. The three models shown are summarized in Table~1.}
\label{fig:plotswithB}
\end{figure*}

\begin{figure}
\centering
\includegraphics[angle=0,width=1.0\columnwidth]{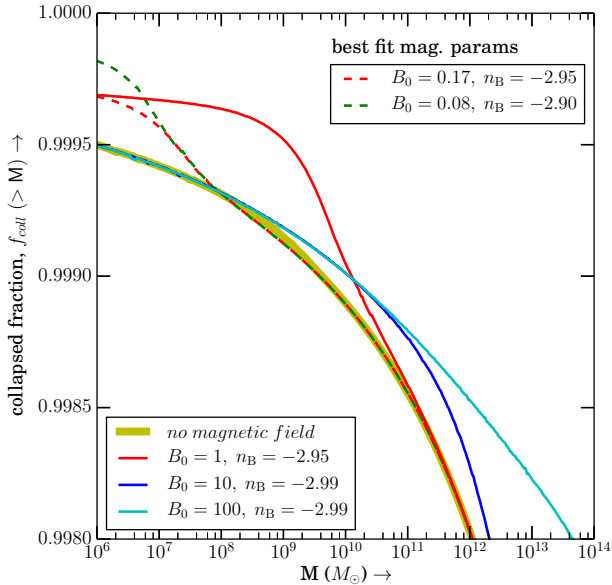}
\caption{Collapsed fraction $f_{\rm coll}(>M)$ of haloes having mass greater than $M$ for $z = 10$. Magnetic field values shown in the figure are in units of nG. 
The dashed lines are for the values of $B_0$ and $n_{\subB}$ which are close to the best-fit values obtained from reionization constraints.}
\label{fig:massfrac}
\end{figure}

\section{Results}

\subsection{Free parameters}

\begin{figure*}
\centering
\includegraphics[angle=0,width=0.99\textwidth]{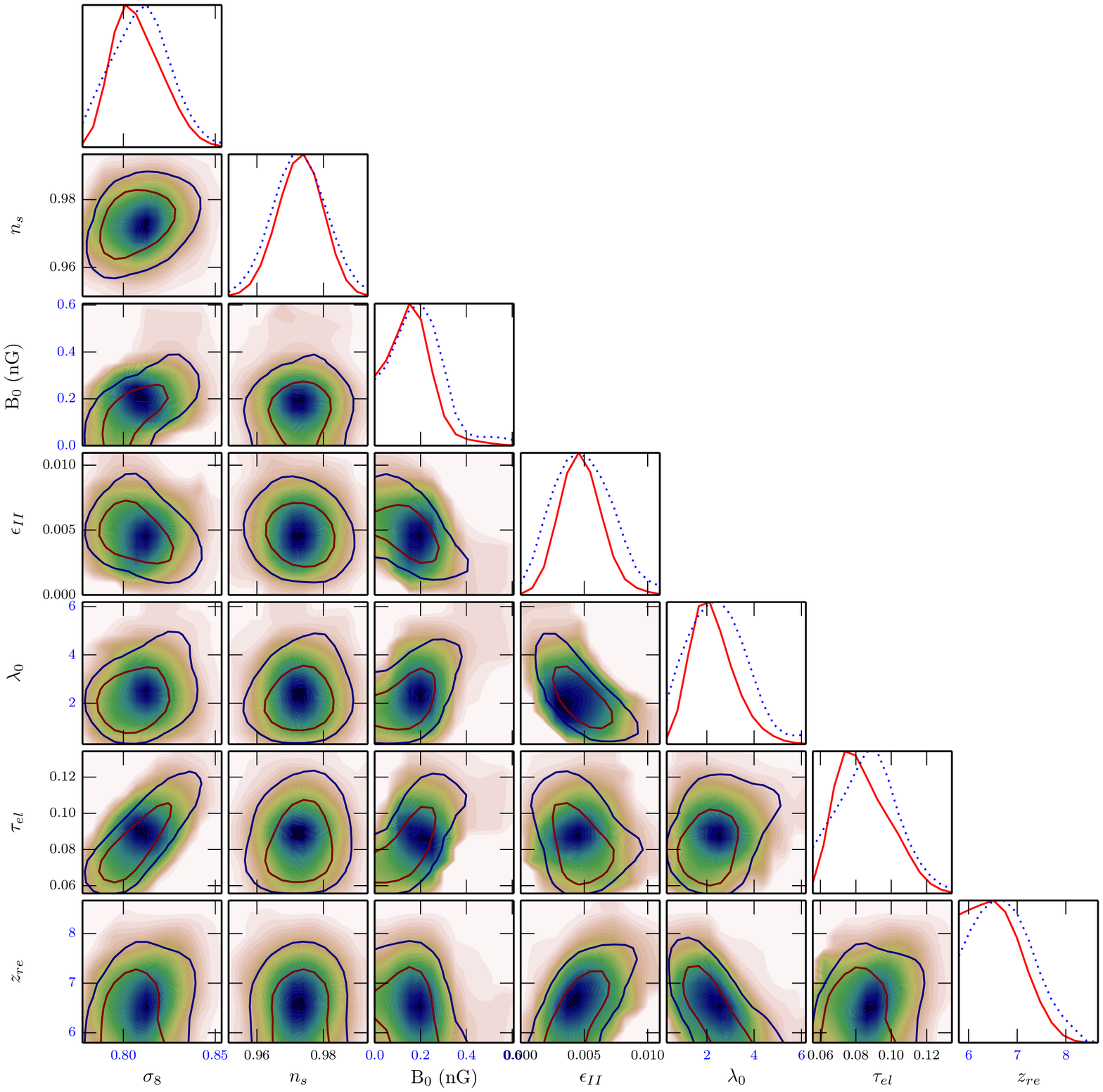}
\caption{Likelihood constraints for $n_{\subB} = -2.95$. The contours are drawn at 68\% and 95\% confidence level. The solid and the dashed curves in the diagonal plots 
          are respectively, the marginalised posterior probability distribution and normalized mean likelihoods of the corresponding parameter.}
\label{fig:fivep-2.99_triplot}
\end{figure*}

\begin{figure*}
\centering
\includegraphics[angle=0,width=0.99\textwidth]{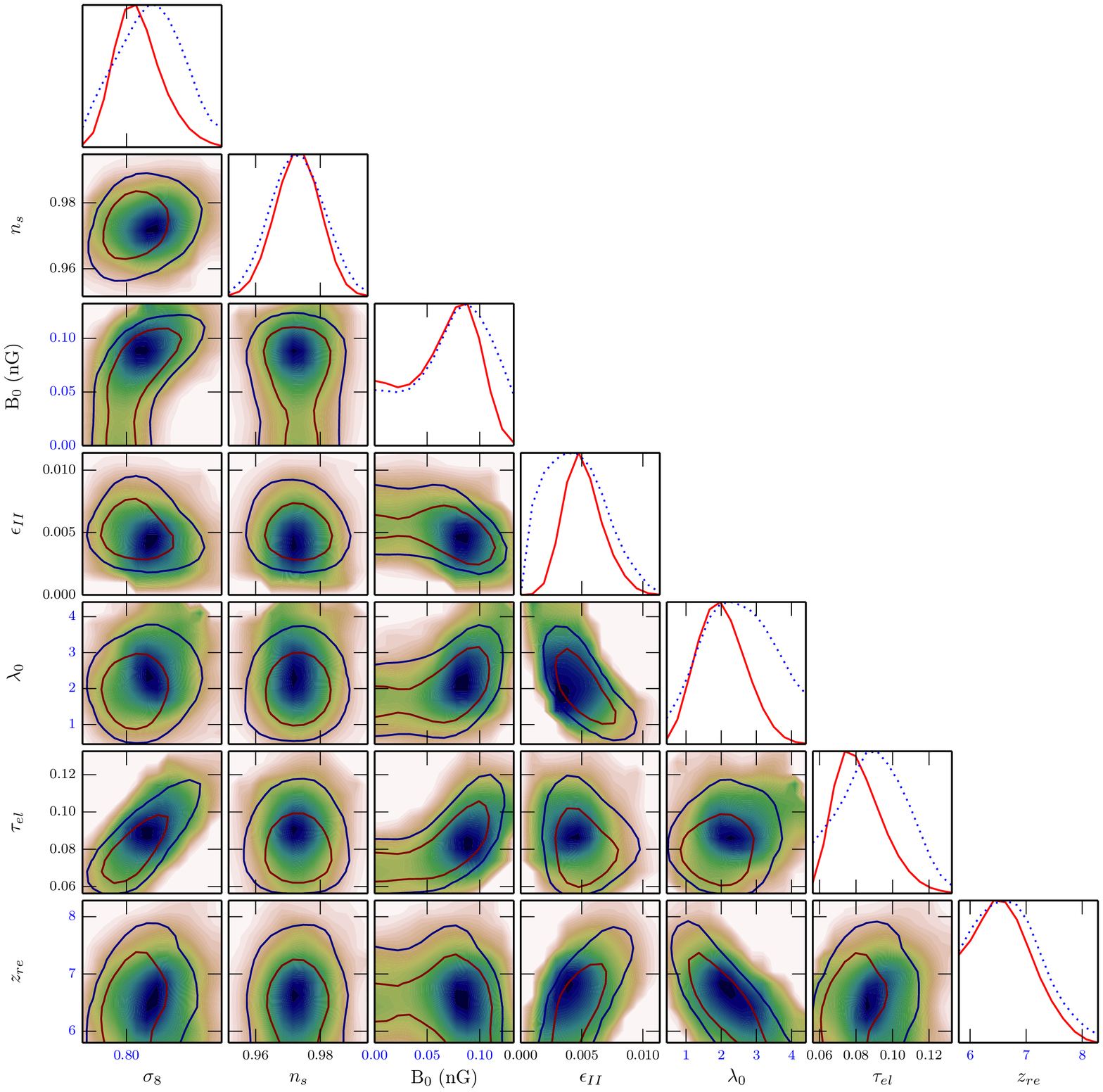}
\caption{Likelihood constraints for $n_{\subB} = -2.90$. The contours are drawn at 68\% and 95\% confidence level. The solid and the dashed curves in the diagonal plots 
          are respectively, the marginalised posterior probability distribution and normalized mean likelihoods of the corresponding parameter.}
\label{fig:fivep_triplot}
\end{figure*}

\begin{figure*}
\centering
\includegraphics[angle=0,width=0.99\textwidth]{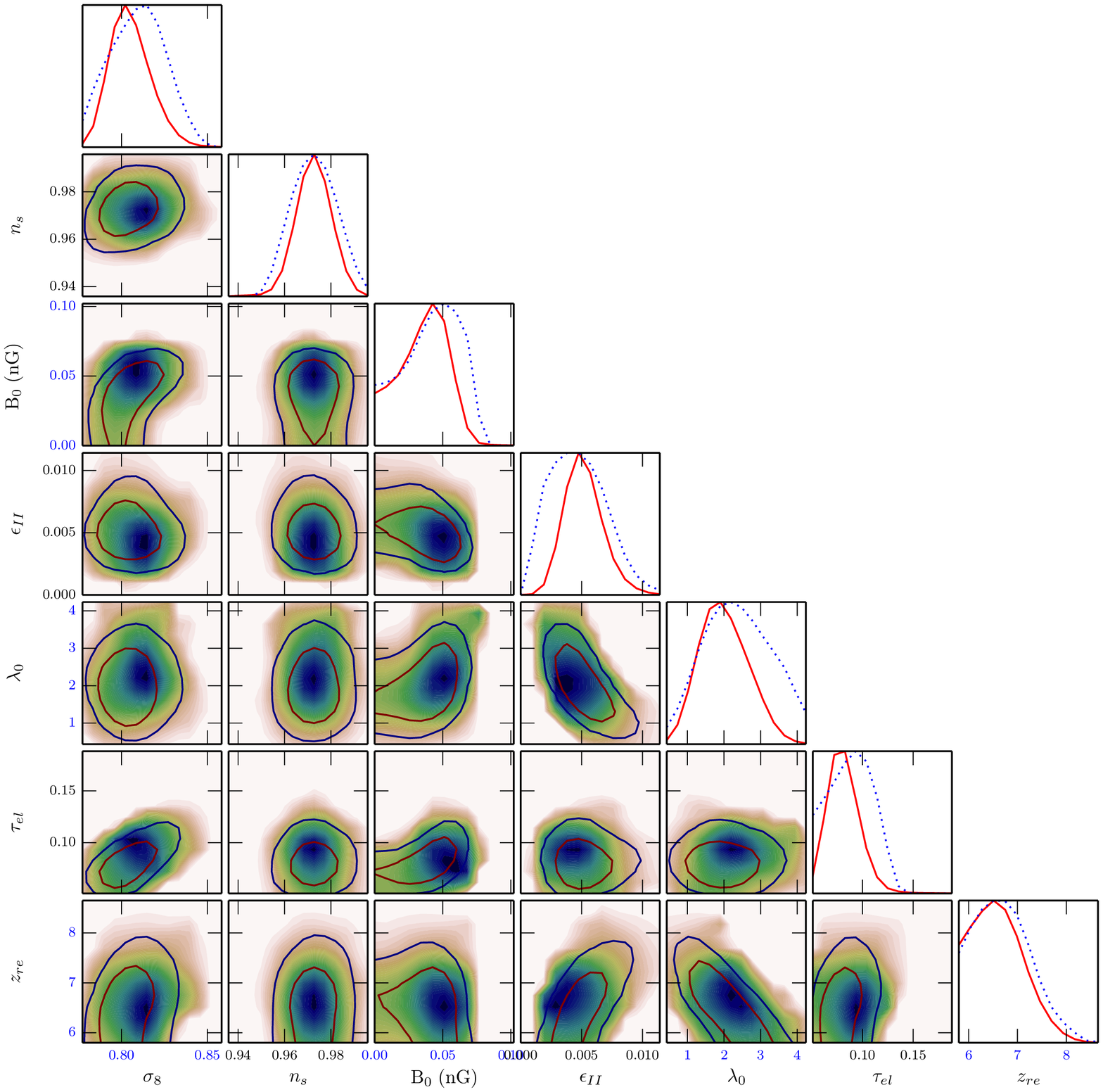}
\caption{Likelihood constraints for $n_{\subB} = -2.85$. The contours are drawn at 68\% and 95\% confidence level. The solid and the dashed curves in the diagonal plots 
          are respectively, the marginalised posterior probability distribution and normalized mean likelihoods of the corresponding parameter.}
\label{fig:fivep-2.90_triplot}
\end{figure*}

In this section, we present the results of our calculation. We 
work with a flat $\Lambda$CDM cosmological model defined by five parameters
$\Omega_m, \Omega_b, h, \sigma_8, n_s$. This has to be supplemented
by the two parameters related to reionization, the stellar efficiency
parameter $\epsilon_{\rm II}$ and the normalization of the
mean free path $\lambda_0$. For models with magnetic field, we need
two more free parameters $B_0$ and $n_{\subB}$ to complete the set.

It has been shown that among the cosmological parameters, the
two which affect the reionization history most
significantly are the ones related to the dark matter fluctuation
power spectrum, i.e. $\sigma_8, n_s$. 
In fact, even when the three parameters $\Omega_m, \Omega_b, h$ are kept fixed, it is found that the best-fit values and the error-bars on the parameters $\epsilon_{\rm II}$ and $\lambda_0$ remain almost identical to the case when all the parameters are allowed to vary \cite{2012MNRAS.419.1480M}. 
For the rest of the paper, we  thus fix the
values of these parameters to their WMAP9 best-fit values
$\Omega_m = 0.279, \Omega_b = 0.046, h=0.70$ \cite{2013ApJS..208...20B}.\footnote {One may argue that it is more reasonable to use priors from more recent experiments like  
Planck \cite{2013arXiv1303.5076P}. We use the WMAP9 priors mainly because the data and likelihood analysis used in section 3.3 are from WMAP9. Using data from \citeN{2013arXiv1303.5076P} may not alter the constraints obtained in this work as most of the parameters have almost similar values (the value of $h$ is smaller for Planck than WMAP, however the combination  $\Omega_m h^2$ remains almost similar, and hence the effect on reionization is not that drastic). It might be interesting to revisit this problem once the Planck polarization measurements are released.}

\subsection{Effect of magnetic fields on reionization history}

We first discuss the effects of non-zero magnetic field 
on the reionization history. In addition to fixing $\Omega_m, \Omega_b, h$ as mentioned above, we fix the values of
the cosmological parameters $\sigma_8 = 0.821, n_s = 0.972$ to their
best-fit values, and also fix $n_{\subB} = -2.9$. The arguments presented in this section would hold equally well for any other value of $n_{\subB}$ close to $-3$. It is sufficient to
vary the efficiency parameter
$\epsilon_{\rm II}$ and the magnetic field $B_0$ to understand the effect.
The models and the parameters considered in this subsection are summarized in Table~\ref{tab:reion_param}.

\begin{table}
                \begin{tabular}{l|c|c|c}
\hline
Parameter     & Model A  & Model B & Model C \\
\hline
$B_0$  (nG)        &   0.0        &   0.0      &   0.08   \\
$\epsilon_{\rm II}$      & 0.006  &   0.0175 &   0.006 \\
\hline               
                \end{tabular}
        \caption{Models used for discussing the effect of $B_0$ on reionization history.}
        \label{tab:reion_param}
\end{table}

We plot the evolution of the electron scattering optical depth $\tau_{\rm el}$, the volume filling factor
of ionized regions $Q_{\rm HII}$ and the hydrogen photoionization rate $\Gamma_{\rm HI}$ in Figure \ref{fig:plotswithB}. 
We also compare the models with relevant observational data, i.e., with the WMAP9 constraints on $\tau_{\rm el}$ \cite{2013ApJS..208...20B} and the measurements of $\Gamma_{\rm HI}$ from Ly$\alpha$ forest data \cite{2011MNRAS.412.1926W,2013MNRAS.436.1023B}.
Consider first model A, which has no magnetic field. The value of $\epsilon_{\rm II}$ is chosen such that
it is consistent with the upper limits of $\Gamma_{\rm HI}$ measurements at $z = 6$ (right hand panel), i.e., this
is the largest $\epsilon_{\rm II}$ consistent with the Ly$\alpha$ forest data. The 
left hand panel shows that this model underpredicts the value of $\tau_{\rm el}$
given  by the  WMAP9 observations. 
If we try to match the $\tau_{\rm el}$ constraints without introducing any additional physics, 
we have to increase the value of $\epsilon_{\rm II}$. Model B represents such a scenario where $\epsilon_{\rm II}$
has been increased by a factor of $\sim 3$ compared to A. Now the match with $\tau_{\rm el}$
is quite good, however, this model overpredicts the $\Gamma_{\rm HI}$ at $z=6$ by a large amount\footnote{Of course, the tension between these two data sets can be reconciled if it is found that the value of $\tau_{\rm el}$ inferred by WMAP is higher than the true value, as is evident from the latest Planck results \cite{2015arxiv150201589P}}.

We now introduce a non-zero magnetic field $B_0 = 0.08$~nG (with $n_{\subB} = -2.9$) in our model. It is possible to repeat the following exercise with any other value of $n_{\subB}$ by choosing an appropriate $B_0$. The efficiency parameter in model~C is kept identical to model~A. We now see that 
the model is able to match both the $\Gamma_{\rm HI}$ data
at $z=6$ and the $\tau_{\rm el}$ constraints. This shows that magnetic field can be useful in relieving the tension
between the CMB and QSO absorption line data sets. The main reason for this is that presence of magnetic field allows for early formation of galaxies and hence can drive early reionization. However, as reionization progresses, the radiative feedback effects become more important and regulates the formation of ionizing sources at $z \sim 10$. This leads to a feedback-regulated, extended and photon-starved reionization which is required for good match with the data \cite{2007MNRAS.382..325B,2011MNRAS.413.1569M,2012MNRAS.419.1480M}\footnote{There are alternate ways to achieve a reionization model which is consistent with all the data sets, e.g., by introducing star formation within minihaloes \cite{2008MNRAS.385L..58C} or by including a population of metal-free stars at high redshifts \cite{2006MNRAS.371L..55C}.}.
It also follows  that this set of observations
can also be used to put constraints  on $B_0$ as too high a value of $B_0$ would lead to reionization too early and would violate the $\tau_{\rm el}$ bounds.

In Figure~3, we show the collapsed fraction for different values
of $B_0$ and the spectral index $n_{\subB}$. This figure allows us to understand 
the results of the current  and the next section where we  present detailed
 multi-parameter analysis. 
As noted above, the main impact of the primordial magnetic fields is to enhance
density fluctuations at small scales (Figure~1). 
Figure~3 shows the impact of this addition to the  collapsed fraction
as a function of mass. The collapsed fraction is seen to   be a 
 sensitive and complex  function of the parameter associated with 
primordial magnetic fields.

We could understand this dependence as follows. 
For the magnetic field-induced density perturbations, the mass dispersion 
at a given scale: $\sigma(M) \propto B_0^2 (n_{\subB} + 3)$ \cite{2014ApJ...786..142V}. Also 
$\sigma(M) \propto M^{-2/3}$ above the magnetic field Jeans' scale which
is a sharper fall as compared to the $\Lambda$CDM model in the 
relevant mass range. This also means that the collapsed fraction 
is dominated by a small range of mass scales around the magnetic field Jeans' 
scale (see e.g. \citeNP{2014ApJ...786..142V}, Figure~3).   In the Press-Schechter formalism used to compute the collapsed fraction, the fraction increases
with  $\sigma(M)$  and  could 
be  exponentially sensitive to the mass dispersion. 
Therefore, we expect an increase in the 
mass fraction as $n_{\subB}$ is increased for a fixed $B_0$. A change in the 
value of $B_0$ results in two distinct effects: (i) $\sigma(M)$ increases 
which tends to increase the collapsed fraction (ii) the magnetic Jeans' length
also increases (for details see discussion in section~2.2) which tends to 
decrease the collapsed fraction below the magnetic field Jeans' mass. The net
effect of increasing the value of $B_0$ is to shift the collapsed fraction to 
larger masses while decreasing the fraction at smaller masses, as seen in Figure~3.

\subsection{Constraints on $B_0$}

In this section, we present 
results related to the constraints on $B_0$ based on  detailed multi-parameter  MCMC analysis.
We have modified the publicly available code COSMOMC\footnote{\tt http://cosmologist.info/cosmomc} \cite{2002PhRvD..66j3511L} to account for generic reionization histories \cite{2011MNRAS.413.1569M,2011PhRvD..84l3522P} and the effect of magnetic field on the matter power spectrum.
We take three different values of $n_{\subB} = -2.95, -2.90, -2.85$ and for each case
we vary five parameters, namely, 
$\sigma_8, n_s, B_0, \epsilon_{\rm II}, \lambda_0$, keeping all the rest of the
parameters fixed. 
We constrain these parameters using (i) the 
WMAP9 data on angular power spectra corresponding to the temperature auto-correlation (TT), the $E$-mode polarization autocorrelation (EE) and the TE cross-correlation \cite{2013ApJS..208...20B}, 
(ii) the photoionization rate in the IGM inferred from the Ly$\alpha$ forest
at $z \leq 6$ \cite{2011MNRAS.412.1926W,2013MNRAS.436.1023B} and (iii) the redshift distribution of Lyman-limit
systems at $z < 6$ \cite{2010ApJ...721.1448S}.

\begin{table*}
                \begin{tabular}{l|c|c|c|c|c|c|c|c|c}
Parameter     &\multicolumn{3}{c}{$n_{\subB} = -2.95$}  & \multicolumn{3}{c}{$n_{\subB} = -2.90$} & \multicolumn{3}{c}{$n_{\subB} = -2.85$} \\
\hline
              & Best-fit & Mean & 95\% c.l. & Best-fit & Mean & 95\% c.l.  & Best-fit & Mean & 95\% c.l. \\
\hline
$\sigma_8$          &  0.811  & 0.808     & [0.786, 0.837]  &  0.810 &0.806     & [0.787, 0.838]  &  0.809  &0.805    & [0.787, 0.835]  \\
$n_s$              &   0.972 & 0.973      & [0.959, 0.986]  &  0.974 &0.973     & [0.960, 0.985]  &  0.973 &0.973     & [0.960, 0.968]  \\
$B_0$  (nG)        &   0.189  & 0.159     & $ < 0.358$     &   0.081  &0.065    & $ < 0.120$    &   0.048  &0.036    & $ < 0.059$           \\
$\epsilon_{\rm II}$      & 0.0042 & 0.0048   &[0.0022, 0.0090] &   0.0051 &0.0052   &[0.0028, 0.0086]&   0.0043 &0.0052   &[0.0030, 0.0087]\\
$\lambda_{0}$ &        2.337  & 2.273      &[0.918 , 4.541 ] &   1.817 &2.021    &[0.944, 4.093]&   2.233 &2.008    &[0.901, 3.721]\\
\hline               
$\tau_{\rm el}$        &   0.089  & 0.084   & [0.063 , 0.118 ]&  0.086 & 0.082  & [0.063, 0.119]  &  0.085 &0.082  & [0.063, 0.119]  \\
$z_{\rm re}$         &       6.400 & 6.579  &[5.800 , 7.800 ]&    6.800 & 6.628    & [5.800, 7.600]&    6.400  &6.624   & [5.800, 7.800]\\
\hline               
$\chi^2$        &   3781.18     & --    & --       &  3781.23 &   --     &--       &  3781.18 &   --     & --       \\
                \end{tabular}
        \caption{Parameter constraints}
        \label{tab:parameters}
\end{table*}

We would like to mention here that, we tried case $n_{\subB} = -2.99$ too but it turned out 
that the effect of magnetic field on the mass function of collapsed haloes is negligible in the mass scales relevant for reionization. The reason is that the magnetic field-induced mass fluctuations have the dependence $\sigma(M) \propto B_0^2 (n_{\subB} + 3)$, and hence one requires very high value of $B_0$ to obtain any significant effect when $n_{\subB} \to -3$.
The collapsed mass fraction $f_{\rm coll}(>M)$ for $B_0 \sim 1$ nG is almost same as the non-magnetic case for 
$M \lesssim 10^{12}$ M$_{\odot}$, see Figure~3, thus implying that no significant effect on reionization history can be expected.
If we try to increase the collapse mass fraction by making magnetic field value very high, 
the effect starts to show at very large scales (e.g., $B_0 \sim 100$ nG shows significant effects only for $M \gtrsim 10^{12}$ M$_{\odot}$). The reason for 
this is that magnetic Jeans cut-off scale is roughly proportional to the magnetic field strength 
($k_J \propto B_{0}^{-1}$). Since the contribution of high mass haloes to the total ionizing photon budget is negligible, 
they hardly make any difference to the reionization history. Consequently the magnetic field 
seems to play no role for $n_{\subB} = -2.99$. This also suggests that the reionization starts to become insensitive to 
magnetic field effects as $n_{\subB} \rightarrow -3$.

We show the results for three fixed values of $n_{\subB}$: \fig{fig:fivep-2.99_triplot}
shows the likelihood contours for $n_{\subB} = -2.95$ (which is almost scale-free).
The contours for  $n_{\subB} = -2.90$ are shown in \fig{fig:fivep_triplot} and the ones
for  $n_{\subB} = -2.85$ are shown in \fig{fig:fivep-2.90_triplot}. The 
results of  our analysis
are summarized in Table \ref{tab:parameters}. Note that in these figures and the tables, the parameter $z_{re}$ 
refers to the redshift when the reionization is 99\% complete as given by our detailed model of reionization.

We can understand our main results   based on the discussion in the previous subsection. The upper
limit on $B_0$ lies in the range $0.36\hbox{--}0.06$~nG and this limit 
decreases as $n_{\subB}$ is increased in the range 
 $n_{\subB} = -2.95$ to $-2.85$, as we expect from the discussion in the 
previous subsection.  We note from the figures that $B_0$ is anti-correlated with the two parameters $\epsilon_{\rm II}$ and $\lambda_0$. This is expected as a decrease in $\epsilon_{\rm II}$ which results in inefficient production of ionizing photons which can be compensated by an increase in $B_0$ leading to early halo formation. Similarly a decrease in $\lambda_0$ leads to smaller values of the photoionization rate which too can be compensated by a higher $B_0$. It is also not surprising that $B_0$ is strongly correlated with the derived parameter $\tau_{\rm el}$, as larger values of $B_0$ leads to early reionization and hence larger $\tau_{\rm el}$.

Our analysis shows  (Table \ref{tab:parameters}) that the 
best-fit value of $B_0$ is always non-zero, i.e., one obtains a better match to the data when a non-zero magnetic field is included in the reionization model. However, it is not possible to rule out the non-magnetic cases within 2-$\sigma$ limits. We also find that the
inferred values of the cosmological parameters
$\sigma_8$ and $n_s$ are quite close to 
what obtained by the WMAP9 team \cite{2013ApJS..208...20B}.

We discuss some of the caveats for our analysis. 
The magnetic Jeans' scale play a crucial role in our analysis. The fact that the magnetic Jeans' scale is not a very well defined 
quantity like the thermal Jeans' scale as it indicates the breakdown of linear analysis in a magnetized fluid (\citeNP{1996ApJ...468...28K}), there 
is an inherent leverage in our analysis. However, as we have discussed in the section~2.1, we take a conservative approach by taking a sharp 
cut-off at magnetic Jeans' scale $k_J$, the constraints on $B_0$ would be even tighter if we use a gradual cut-off.
In the reionization model studied  here,  the assumption of
a redshift-independent $\epsilon_{\rm II}$ is used.
 Some of the constraints on $B_0$
might change if this assumption is relaxed. For example, if the effective $\epsilon_{\rm II}$
is allowed to increase at high redshifts (e.g., introducing a population
of metal-free stars and/or X-ray ionizing sources), then the 
upper-limit on $B_0$ would become much tighter. On the other hand, if there are
reasons to believe that $\epsilon_{\rm II}$ decreases at high-$z$ (e.g., 
if the escape fraction of photons is less at early epochs), then
one would need higher values of $B_0$ to explain the data and hence the
upper-limit on $B_0$ would be weaker. Interestingly, one may end up putting
a lower-limit on $B_0$ in such a case. Similarly, the constraints on $B_0$ could be degenerate with the effects of feedback. In this work, we have used a simple model of radiative feedback based on Jeans prescription, while the actual situations could be more complex. It would be interesting to check how the constraints change with different prescriptions for feedback. 
The other crucial assumption in our work is that we ignore the possibility of molecular cooling in haloes. If such an effect is introduced, it will allow star formation in minihaloes and would thus favour early reionization. In that case we would be able to put much tighter limits on $B_0$.
We shall explore such interesting
possibilities in a later work.

\section{Discussion}

We studied the possible role primordial magnetic fields might play in 
explaining the reionization history of the universe in the redshift range
$z \simeq 6\hbox{--}10$. 
 These  fields enhance  the  power in the
dark matter density fluctuations at small scales thus allowing early structure formation.
Our main results are: (i) a non-zero  $B_0$ helps in relieving the
tension between CMB and quasar absorption line data sets in the photon-starved
reionization scenario by enabling early structure formation, and (ii) the data sets 
can be useful in putting a upper-limit on $B_0$, we obtain $B_0 < 0.358, 0.120, 0.059$ nG (95 \% c.l.) for 
$n_{\subB} = -2.95, -2.9, -2.85$, respectively. 

Many cosmological observables have been analysed to constrain the 
amplitude and the spectral index of the magnetic field power spectrum: 
CMB observations, early structure formation, weak gravitational lensing, 
Lyman-$\alpha$ data, etc.; these considerations put  upper bounds on $B_0$ in  the range 
$0.3\hbox{--}1$~nG (e.g. \citeNP{2012PhRvL.108w1301T},  \citeNP{2010PhRvD..82h3005K}, \citeNP{2012PhRvD..86d3510S}, \citeNP{2013ApJ...762...15P}, \citeNP{2012ApJ...748...27P}, \citeNP{2004PhRvD..70d3011L}, \citeNP{2004PhRvD..69f3006C}).  
Bounds obtained from Big Bang Nucleosynthesis constraints give $B_0 \lesssim 10^{-7}$~G \cite{1999PhRvD..59l3002S}.
Earlier constraints on primordial magnetic fields coming from the study of ionization history of the post-recombination 
universe (\citeNP{2006MNRAS.368..965T}, \citeNP{2008PhRvD..78h3005S}, \citeNP{2011MNRAS.418L.143S}) also put the upper bound on $B_0$ in the reange $0.7\hbox{--}5$~nG.
Our results are consistent with all these constraints. 
Interestingly, even  magnetic fields of smaller magnitude $\simeq 0.1$~nG can have an appreciable and potentially detectable 
impact on the reionization history. In the future, one can possibly improve these bounds by understanding some of the physical processes related to reionization (e.g., feedback) through detailed modelling.

\bibliographystyle{mnras}
\bibliography{mnrasmnemonic,ms}

\end{document}